\documentclass[12pt]{article}
\usepackage{amsmath,amssymb,graphicx}
\newcommand{\be}{\begin{equation}}
\newcommand{\bea}{\begin{eqnarray}}
\newcommand{\eea}{\end{eqnarray}}
\newcommand{\ba}{\begin{array}}
\newcommand{\ea}{\end{array}}
\newcommand{\ee}{\end{equation}}

\begin{document}

\begin{titlepage}

\vspace*{5mm}%
\begin{center}

{{\Large {\bf Holographic Cusped Wilson loops in q-deformed $AdS_5\times S^5$ Spacetime}}}

\vspace*{15mm} \vspace*{1mm} {Nan Bai$ \footnote{ bainan@ihep.ac.cn}$, Hui-Huang Chen$\footnote{chenhh@ihep.ac.cn}$, Jun-Bao Wu$\footnote{wujb@ihep.ac.cn}$}

 \vspace*{2cm}

{\it
Institute of High Energy Physics, and Theoretical Physics Center for Science\\
Facilities,\\
Chinese Academy of Sciences, 19B Yuquan Road, Beijing 100049, P. R. China}

\vspace*{.4cm}

\vspace*{2cm}
\end{center}
\begin{abstract}
In this paper, minimal surface in $q$-deformed $AdS_5\times S^5$ with boundary a cusp is studied in detail.
This minimal surface is dual to cusped Wilson loop in the dual field theory.
We found that the area of the minimal surface has both logarithmic squared divergence and logarithmic divergence.
The logarithmic squared divergence can not be removed by either Legendre transformation or the usual geometric substraction.
We further make analytic continuation to the Minkowski signature 
and take the limit such that the two edges of the cusp become light-like
and extract anomalous dimension from the coefficient of the logarithmic divergence. This anomalous dimension goes back smoothly
to the results in the undeformed case when we take the limit that the deformation parameter goes to zero.
\end{abstract}
\end{titlepage}

\section{Introduction}
Integrability \cite{Beisert:2010jr} and localization \cite{Pestun:2007rz} make us now be able to compute some important quantities in ${\cal N}=4$ super Yang-Mills theory
 as non-trivial functions of 't Hooft coupling $\lambda$ and the rank of the gauge group $N$ \footnote{Integrability is mainly established
 in the large $N$ limit.}. These computations lead to results at strong coupling  and give non-trivial test on the famous $AdS/CFT$
 correspondence \cite{Maldacena:1997re, Gubser:1998bc, Witten:1998qj}. Cusp anomalous dimension $f(\lambda)$ is among these interesting quantities and its value at finite $\lambda$ in the planar limit can be computed using this powerful integrability method \cite{Beisert:2006ez, Freyhult:2010kc}. This function appears as a cusp anomaly of a light-like Wilson loop
 \cite{Polyakov:1980ca, Brandt:1981kf}. It also appears as the coefficient in front of $\log S$ of the anomalous dimension of large spin twist-two operator (here $S$ is the spin of this operator)\cite{Gross:1974cs, Georgi:1951sr}. The fact that these two approaches give the same function $f(\lambda)$ was proved
 in the perturbative gauge theory in \cite{Korchemsky:1988si, Korchemsky:1992xv, Bassetto:1993xd}.

Both approaches for the cusp anomalous dimension have dual descriptions in the gravity side of gauge/string duality. The twist-two operator is dual to folded spinning strings in $AdS_5$ found by Gubser-Klebanov-Polyakov (GKP) \cite{Gubser:2002tv}. The anomalous dimension of the operator can be obtained from the energy of the semi-classical string. The Wilson loop is dual to an open F-string in $AdS_5$, and the contour of the  Wilson loop is just the boundary of the  F-string  worldsheet \cite{Rey, Mal98}. The holographical dual of cusped light-like Wilson loop was studied in detail in \cite{c3} (see also \cite{Makeenko:2002qe}) by performing a nontrivial analytic continuation of the F-string solution dual to cusped Wilson loop
in Euclidean space in \cite{c1}. The cusp anomalous dimension obtained from the open F-string solution coincides with the results from the closed string solution obtained in \cite{Gubser:2002tv}. In \cite{c3}, this was taken as evidence that \cite{Gubser:2002tv} made  the correct identification for  string theory dual of the twist-two operators. In \cite{Kruczenski:2007cy}, the scaling limits of the above closed string solution and open string solution
was shown to be equivalent through a analytic continuation and an $AdS_5$ isometry rotation. This explained, in the gravity side, why this two approaches
give the same results for the anomalous dimension. This can be thought as a kind of open-closed duality in $AdS$ background.

It is obviously of great value to search for integrable structure in $AdS/CFT$ correspondence with less supersymmetries.
Such examples are very rare. Orbifolds \cite{Ideguchi:2004wm}-\cite{Solovyov:2007pw}, $\beta$- and $\gamma$-deformations
\cite{Roiban:2003dw}-\cite{Frolov:2005dj}, and adding suitable fundamental matters \cite{Chen:2004mu}-\cite{Mann:2006rh}
are almost the only known examples where the four-dimensional field theories are still the usual local gauge theories and the integrability in the planar limit is preserved \cite{Zoubos:2010kh}.
Many other four-dimensional theories and their gravity duals are not integrable. Instead they display chaotic behaviors \cite{Zayas:2010fs}-\cite{Chervonyi:2013eja}.
In the gravity dual of the orbifolds and $\beta$($\gamma$)-deformation examples, the $AdS_5$ part is untouched. For the first case, the five-sphere is replaced
by its orbifolds. And for the second case, the metric on $S^5$ is deformed with other background fields turned on. Since the $AdS_5$ part of the metric is unchanged and the NS-NS field has vanishing components in $AdS_5$ part for both cases, the computations for both GKP folded string in $AdS_5$
and F-string dual to cusped Wilson loops are not changed. The above mentioned open-closed duality in $AdS$ space is preserved in a trivial manner. We also notice that this open-closed duality were also found for string theory on $AdS_3\times S^3\times M_4$ with both NS-NS and RR three-form fluxes \cite{David:2014qta}.

It is then quite interesting to search for integrable models with a gravity dual involving more complicated geometry replacing the $AdS$ part.
Remarkably, one of such integrable deformations on the  worldsheet theory was constructed in \cite{Delduc:2013qra}. Many aspects of such deformation was already studied in \cite{Arutyunov:2013ega}-\cite{Ahn:2014iia}. The background was called $q$-deformed $AdS_5\times S^5$ \footnote{Some people chose the name $\eta$-deformation.}. The field theory dual of string theory on this background is still unclear. People hope that the studies of various aspects in the
string theory side can give us some hints on the possible dual field theory. Many classical string solutions in this background were studied in detail
in \cite{Ahn:2014aqa, Arutyunov:2014cda, Kameyama:2014vma,Banerjee:2014bca, FrolovRoiban,c4}. There are already several interesting features for the classical strings which are different from the case without deformations. The GKP spinning string solutions found in \cite{Kameyama:2014vma,FrolovRoiban}
can not be smoothly connected with the original solutions in \cite{Gubser:2002tv} when we take the limit that the deformation parameter goes to zero.
And the energy $E$ and spin $S$ of these spinning strings  will not have the relation $E-S\sim f(\lambda)\log S$ in the large $S$ limit.
Another interesting result \cite{c4} is that the open F-string solution with boundary a circle has finite area without peforming geometric substraction or Legendre transformation which was used for the undeformed case, though there are divergences in the action when the boundary is a straight line \cite{Kameyama:2014vma}. Here the deformation parameter plays the role of a UV regularization \cite{c4}.

The above features led us to the study of the F-string solution  with boundary a cusp in $q$-deformed $AdS_5\times S^5$.
We also consider the case when there is a jump in the deformed $S^5$ at the cusp. The solution was found by computing the conserved charges
from the symmetry of the system. We find the area of the worldsheet has behavior different from both the case with circle as boundary and the
holographic dual of cusped Wilson loops in the undeformed case. The area has logarithmic squared divergence, in additional with the logarithmic divergence.
The logarithmic squared divergence is softer than the linear divergence in the undeformed case. However the UV regularization provided by the deformation parameter did not change it to be finite. We then turn to the attempts for the  renormalization of the area. Two usually-used methods, Legendre transformation and geometric substraction are considered. We found that neither of them can remove the logarithmic squared divergence, and these two methods are no longer equivalent
to each other.  Finally, by continuation to the Minkowski signature and subtracting the logarithmic squared divergence by hand, we computed the cusp anomalous dimension  for the deformed case. We find that this result can be smoothly connected with the result in the undeformed case when we take the limit that
the deformation parameter tends to zero.

The remaining part of this paper is as follows: in the next section, we will find the F-string solution in $q$-deformed $AdS_5\times S^5$ dual to cusped Wilson loops. 
The cusp anomalous dimension will be extracted from the cusped Wilson loops in section 3. Final section is devoted to conclusion and discussions.

\section{F-string solution dual to cusped Wilson loop}
\subsection{$q$-deformed $AdS_5\times S^5$ }
In \cite{Delduc:2013qra}, an integrable deformation of type IIB superstring theory on $AdS_5\times S^5$ was constructed.
From this, the string frame metric and B-field for this string background was given in \cite{Arutyunov:2013ega}.
Later a new coordinate system was introduced in \cite{Kameyama:2014vma} which was inspired by studies of GKP (Gubser-Klebanov-Polyakov) strings \cite{Gubser:2002tv} in $q$-deformed $AdS_5$. A related Poincare-like coordinate system for $q$-deformed $AdS_5$  was introduced in \cite{c4}.  This will be our starting point.
So now  we list the results of metric and B-field in this Poincare-like coordinates.
The metric for the $q$-deformed AdS part in Poincare coordinates is,
\bea
ds^2&=&\sqrt{1+C^2}R^2\left[\frac{dy^2+dr^2}{y^2+C^2(y^2+r^2)}+\frac{C^2(ydy+rdr)^2}{y^2(y^2+C^2(y^2+r^2))}\right.\nonumber\\
&&\left.+\frac{(y^2+C^2(y^2+r^2))r^2}{(y^2+C^2(y^2+r^2))^2+C^2r^4\sin^2\zeta}(d\zeta^2+\cos\zeta^2 d\phi^2)+\frac{r^2\sin^2\zeta d\psi^2}{y^2+C^2(y^2+r^2)}\right].\label{deformedads}
\eea
The metric for the $q$-deformed $S^5$ part is,
\bea
ds^2&=&\sqrt{1+C^2}R^2\left[\cos^2\gamma d\theta^2+\frac{d\gamma^2}{1+C^2\cos^2\gamma}\right.\nonumber\\
&&\left.+\frac{(1+C^2\cos^2\gamma)\sin^2\gamma}{(1+C^2\cos^2\gamma)^2+C^2\sin^4\gamma\sin^2\xi}(d\xi^2+\cos^2\xi d\phi_1^2)+\frac{\sin^2\gamma\sin^2\xi d\phi_2^2}{1+C^2\cos^2\gamma}\right].
\eea
The action of Wess-Zumino term for the deformed AdS part is,
\bea
\mathcal{L}^{WZ}_{1}=\frac{C\sqrt{1+C^2}R^2}{4\pi\alpha^\prime}\epsilon^{\mu\nu}\frac{r^4\sin 2\zeta\partial_{\mu}\phi\partial_{\nu}\zeta}{\left[C^2r^2+(1+C^2)z^2\right]^2+C^4r^4\sin^2\zeta},
\eea
and for the deformed $S^5$ part is,
\bea
\mathcal{L}^{WZ}_{2}=-\frac{C\sqrt{1+C^2}R^2}{4\pi\alpha^\prime}\epsilon^{\mu\nu}\frac{\sin^4\gamma\sin2\xi}{(1+C^2\cos^2\gamma)^2+C^2\sin^4\gamma\sin^2\xi}
\partial_{\mu}\phi_1\partial_{\nu}\xi.
\eea
It is easy to see that $C$ plays the role of deformation parameter and when we take the limit $C\to 0$, we will go back to the undeformed case.
\subsection{Cusped Wilson loop}

\subsubsection{Loops without a jump in deformed $S^5$}
We now begin our computation of  minimal surface with boundary a cusped  loop.
This minimal surface is the worldsheet of F-string in deformed $AdS_5\times S^5$ dual to a cusped Wilson loop in the dual field theory.
First we study the case with trivial dependence on the coordinates of deformed $S^5$, that is to say that the coordinates of deformed $S^5$
take constant values on the worldsheet.

At the boundary the Wilson loop is put at two lines
  \be r\in [0, \infty), \psi=0, \zeta=\frac{\pi}{2}, \phi=0, \ee
  and \be r\in [0, \infty), \psi=\Omega, \zeta=\frac{\pi}{2}, \phi=0. \ee
 The string worldsheet will extend to the bulk of deformed $AdS_5$. Let us choose $r$ and $\psi$ to be the coordinates of string worldsheet
 and start with the following ansatz
 \be y=y(r, \psi), \zeta=\frac{\pi}{2}, \phi=0,\ee
 the boundary condition is \be
y(r,0)=y(r,\Omega)=0.
\ee
  Taking into account the invariance of the metric in eq.~(\ref{deformedads}) under the scaling transformation
  \be  y \to \lambda y, r\to \lambda r,\ee
we expect the solution for $y(r, \psi)$ should take the form of
\be y(r, \psi)=\frac{r}{f(\psi)}. \ee
And the boundary condition gives now
\be \lim_{\psi\to 0}f(\psi)=\lim_{\psi\to \Omega}f(\psi)=\infty.\ee
One can also check that the Wess-Zumino term in the worldsheet action will not affect the equation of motion
for the ansatz chosen above \footnote{Things will be different if we choose $\zeta=0, \psi=0$ and worldsheet along $y, r, \phi$ directions. In this case,
though WZ term will not contribute the worldsheet action, it does affect the string eqaiton of motion.}.

Substituting the ansatz back into the target space metric in eq.~(\ref{deformedads}), we obtain the induced metric on the worldsheet ,
\be
ds^2_{ind}=R^2\sqrt{1+C^2}\left[\frac{1+f^2}{r^2}dr^2-\frac{2f'}{rf}dr d\psi+\frac{1}{f^2}\frac{(1+C^2)f'^2+f^4}{1+C^2(1+f^2)}d\psi^2\right].
\ee
Then the area of the surface is
\be
A=\sqrt{1+C^2}R^2\int drd\psi\frac{1}{r}\sqrt{\frac{f^2+f^4+f'^2}{1+C^2+C^2f^2}}. \label{f1}
\ee
So the Nambu-Goto action of the string is
\bea
S_{NG}&=&\frac{1}{2\pi\alpha^\prime}A\nonumber\\
&=&\frac{\sqrt{1+C^2}R^2}{2\pi\alpha^\prime}\int drd\psi\frac{1}{r}\sqrt{\frac{f^2+f^4+f'^2}{1+C^2+C^2f^2}}.
\eea

Therefore, finding the minimum surface in the bulk in this case reduces to a one dimension variational problem with the Lagrangian,
\be
{\cal L}=\int d\psi\sqrt{\frac{f^2+f^4+f'^2}{1+C^2+C^2f^2}}.
\ee
We can solve this extreme value problem by making use of the translation invariance in $\psi$ ($L$ does not depends on $\psi$ explicitly), and the corresponding conserved charge is:
\be
E=\frac{1}{\sqrt{1+C^2+C^2f^2}}\frac{f^4+f^2}{\sqrt{f^4+f^2+f'^2}}.
\ee
Due to the symmetry of the system, $f$ will achieve its minimal value $f_0$ at $\psi=\Omega /2$, then we have $\partial_{\psi}f|_{\psi=\Omega/2}=0$. Thus we can also express $E$ in terms of $f_0$,
\be
E=\frac{f_0\sqrt{1+f_0^2}}{\sqrt{1+C^2+C^2f_0^2}}.
\ee
We may work out $f'$ by equating these two expressions of $E$,
\be
\left(\frac{df}{d\psi}\right)^2=\frac{(f^4+f^2)(f^2-f_0^2)(1+f_0^2+f^2+C^2(1+f_0^2+f^2+f_0^2f^2))}{f_0^2(1+f_0^2)(1+C^2+C^2f^2)}. \label{f2}
\ee
From this, we get the relation between $f_0$ and $\Omega$ as
\bea \frac{\Omega}2&=&f_0\sqrt{1+f_0^2}\int_{f_0}^\infty \frac{df}f\nonumber\\
 &\times& \sqrt{\frac{1+C^2+C^2f^2}{(f^2-f_0^2)(1+f^2)(1+f^2+f^2_0+C(1+f_0^2)(1+f^2))}}.\eea
By making the transformation  $f=\sqrt{z^2+f_0^2}$, we get
\bea \frac{\Omega}2&=&f_0\sqrt{1+f_0^2}\int_{0}^\infty dz\nonumber \\
&\times& \frac{\sqrt{1+C^2(1+z^2+f^2_0)}}{(z^2+f_0^2)\sqrt{(1+z^2+f_0^2)(1+2f_0^2+z^2+C^2(1+f_0^2)(1+f_0^2+z^2))}}.\eea
Substituting eq.~(\ref{f2}) back into the Lagrangian , we obtain,
\be
{\cal L}=2\sqrt{1+C^2}R^2\int^{\frac{r}{\epsilon}}_{f_0}df\frac{\sqrt{(1+C^2+f_0^2C^2)f^2(1+f^2)}}{\sqrt{(f^2-f_0^2)(1+C^2+C^2f^2)(f_0^2+(1+C^2+f_0^2C^2)(f^2+1))}},
\ee
where we have imposed an infrared cutoff for $y$,\quad$y>\epsilon$ or $f<r/\epsilon$. We can further  make the transformation  $f=\sqrt{z^2+f_0^2}$ as above and  the
integral becomes,
\be
{\cal L}={\cal L}(r,\epsilon)=\frac{2\sqrt{1+C^2}R^2}{C}\int^{\sqrt{\frac{r^2}{\epsilon^2}-f_0^2}}_0dz\frac{\sqrt{z^2+a}}{\sqrt{(z^2+b)(z^2+c})},
\ee
with $a,b,c$ listed below,
\bea
a&=&1+f_0^2,\\
b&=&\frac{1+C^2+C^2f_0^2}{C^2},\\
c&=&\frac{1+2f_0^2+C^2(1+2f_0^2+f_0^4)}{1+C^2+f_0^2C^2}.
\eea
Here we will give an approximate analysis since a special function solution requires additional constraints for the parameters $f_0$ and $C$ and will not make the result clearer. By making a change of variable $z=1/t$ and
noticing that
\be \sqrt{\frac{r^2}{\epsilon^2}-f_0^2}\approx \frac{r}{\epsilon}, \ee for small $\epsilon$,
 we have
\be
{\cal L}(r,\epsilon)\approx\frac{2\sqrt{1+C^2}R^2}{C}\int^{\infty}_{\epsilon/r}\frac{dt}{t}\sqrt{\frac{1+at^2}{(1+bt^2)(1+ct^2)}}.
\ee
To extract the divergent part of the integral, we  expand the integrand around $t=\epsilon/r$,
\be
{\cal L}(r,\epsilon)\approx\frac{2\sqrt{1+C^2}R^2}{C}\log\frac{r}{\epsilon}+L_{finite},
\ee
Hence, the area can be evaluated as,
\be
S_{NG}\approx\frac{\sqrt{1+C^2}R^2}{2\pi \alpha^\prime C}\log^2\frac{L}{\epsilon}-\frac{1}{2\pi}F(\Omega,C)\log\frac{L}{\epsilon}, \label{ng}
\ee
where $L$ is the cutoff for the length of the two rays of the Wilson loop and the function $F$ comes from the finite part of ${\cal L}(r,\epsilon)$. We find that the area is composed of two kinds of divergences, logarithmic one and logarithmic squared one while for the undeformed case there are linear divergence plus a logarithmic one \cite{c1}.
Unlike the undeformed case where the linear divergence can be casted away by means of Legendre transformation, we can not manage to subtract any divergence in this way in
the deformed background as will demonstrated in the next subsection.

\subsubsection{Loops with a jump in deformed $S^5$ at the cusp}
We continue to study  a cusped loop where the points on the two edges correspond to two different points in deformed $S^5$. In this work, we will only consider the case that  the dual of the  two edges have a relative angle of $\Theta$ only along the $\theta$ direction of deformed $S^5$. For the undeformed case, a complete and analytical solution is given in \cite{c2}. 

Since the cusp is still invariant under the rescaling of $r$,  we consider the following ansatz,
\bea
y(r,\psi)=\frac{r}{f(\psi)},\qquad \zeta=\frac{\pi}{2},\qquad \phi=0,\nonumber\\
\gamma=\xi=\phi_1=\phi_2=0,\qquad\theta=\theta(\psi).
\eea
Therefore the induced metric on the worldsheet turns out to be,
\bea
h_{rr}&=&R^2\sqrt{1+C^2}\cdot\frac{1+f^2}{r^2},\\
h_{r\psi}&=&R^2\sqrt{1+C^2}\cdot\frac{-f'}{rf},\\
h_{\psi\psi}&=&R^2\sqrt{1+C^2}\left[\frac{(1+C^2)f'^2+f^4}{f^2+C^2f^2(1+f^2)}+\theta'^2\right].
\eea
The area becomes,
\be
A={\sqrt{1+C^2}R^2}\int\frac{dr}{r}\int_0^{\Omega} d\psi\sqrt{\frac{f'^2+f^2+f^4}{1+C^2+C^2f^2}+(1+f^2)\theta'^2}. \label{f3}
\ee
We focus on the Lagrangian density,
\be
\mathcal{L}=\sqrt{\frac{f'^2+f^2+f^4}{1+C^2+C^2f^2}+(1+f^2)\theta'^2}.
\ee
It is easy to find two conserved charges of the system, the energy and the canonical momentum conjugate to $\theta$,
\bea
E&=&\frac{1}{\sqrt{1+C^2+C^2f^2}}\frac{f^4+f^2}{\sqrt{f'^2+f^2+f^4+(1+C^2+C^2f^2)(1+f^2)\theta'^2}},\\
J&=&\frac{1+f^2}{\mathcal{L}}\theta'.
\eea
For the convenience of calculation, we introduce two new conserved quantities which are the combinations of $J$ and $E$,
\bea
p&=&\frac{1}{E},\\
q&=&\frac{J}{E}=\frac{1+C^2+C^2f^2}{f^2}\theta'.
\eea
We find immediately $\theta'=qf^2/(1+C^2+C^2f^2)$ and by substituting it into $p$, $f'$ is easily obtained,
\be
\left(\frac{d\psi}{df}\right)^2=\frac{1}{f^2(1+f^2)}\frac{1+C^2+C^2f^2}{f^2((1+f^2)p^2-q^2-C^2)-C^2-1}.
\ee
The extreme value $f_0$ is determined from the condition $\partial_{\psi}f|_{\psi=\Omega/2}=0$ as follows,
\bea
\frac{f_0^2(1+f_0^2)p^2-f_0^2q^2}{1+C^2+C^2f_0^2}=1.
\eea
The relation between $\Omega$ and $f_0$ is
\bea \frac{\Omega}2&=&\int_{f_0}^\infty \frac{df\sqrt{1+C^2+C^2f^2}}{f\sqrt{(1+f^2)(f^2((1+f^2)p^2-q^2-C^2)-C^2-1)}},\eea
as in the previous subsection, we make the transformation $f=\sqrt{f^2_0+z^2}$ and obtain
\be \Omega=2\int_0^\infty \frac{dz \sqrt{1+C^2(1+f_0^2+z^2)}}{(f^2_0+z^2)\sqrt{(1+f^2_0+z^2)(p^2z^2+p^2(1+2f_0^2)-q^2-C^2)}}. \ee
The relaiton between $\Theta$ and $f_0$ is
\bea
\Theta&=&\int_0^\Omega \frac{q f^2}{1+C^2(1+f^2)} d\psi\\
&=&2\int_{f_0}^\infty \frac{qfdf}{\sqrt{(1+C^2(1+f^2))(1+f^2)(f^2((1+f^2)p^2-q^2-C^2)-C^2-1)}}\\
&=&2q\int_0^\infty \frac{dz }{\sqrt{(1+f^2_0+z^2)(1+C^2 (1+f_0^2+z^2))(p^2z^2+p^2(1+2f_0^2)-q^2-C^2)}}.
\eea

The area becomes,
\bea
A&=&{\sqrt{1+C^2}R^2}\int\frac{dr}{r}\int_0^{\Omega} d\psi \mathcal{L}\nonumber\\
&=&{\sqrt{1+C^2}R^2}\int\frac{dr}{r}\int_0^{\Omega} d\psi \frac{pf^2(1+f^2)}{1+C^2+C^2f^2}\nonumber\\
&=&{2\sqrt{1+C^2}R^2}\int\frac{dr}{r}\int_{f_0}^{r/\epsilon} df \frac{pf\sqrt{(1+f^2)}}{\sqrt{1+C^2+C^2f^2}}\frac{1}{\sqrt{f^2((1+f^2)p^2-q^2-C^2)-C^2-1}}\nonumber\\
&=&\frac{2\sqrt{1+C^2}R^2}{ C}\int\frac{dr}{r}\int_0^{\sqrt{\frac{r^2}{\epsilon^2}-f_0^2}} dz\frac{\sqrt{z^2+k_1}}{\sqrt{z^2+k_2}\sqrt{z^2+k_3}},
\eea
where
\bea
k_1&=&f_0^2+1,\\
k_2&=&\frac{1+C^2+C^2f_0^2}{C^2},\\
k_3&=&\frac{p^2(2f_0^2+1)-q^2-C^2}{p^2}.
\eea
We can analyze the integral approximately by using a new variable $t=1/z$,
\bea
&&\int_0^{\sqrt{\frac{r^2}{\epsilon^2}-f_0^2}} dz\frac{\sqrt{z^2+k_1}}{\sqrt{z^2+k_2}\sqrt{z^2+k_3}}\nonumber\\
&\approx&\int^{\infty}_{\frac{\epsilon}{r}}dt\frac{1}{t}\frac{\sqrt{1+k_1t^2}}{\sqrt{(1+k_2t^2)(1+k_3t^2)}}\nonumber\\
&\approx&\log\frac{r}{\epsilon}+\mbox{finite terms}.
\eea
Therefore the area is,
\be
A\approx\frac{\sqrt{1+C^2}R^2}{C}\log^2\frac{L}{\epsilon}-\frac{1}{2\pi}F(\Omega,\Theta,C)\log\frac{L}{\epsilon}.
\ee
We find that the structure of the divergences is the same as the no jump case.
\subsection{Renormalization of the area}
 Let us first recall the story in the undeformed case. When the contour of the Wilson loops is smooth, the bare area of the F-string worldsheet has divergence universally  as ${L}/\epsilon$ \cite{Polyakov:2000ti} where $L$ is the length of the loop and $\epsilon$ is the cut-off as introduced in this paper. This divergence can be removed either via a Legendre transformation \cite{c1} or by a geometric substraction \cite{Mal98}. And these two methods are equivalent to each other. For the case with a cusp, beside this ${L}/\epsilon$ term, there is a subleading divergence term growing as $\log(L/\epsilon)$.  The leading divergence can be removed by either
 of the two methods, and the subleading $\log(L/\epsilon)$ term will remain there. This is consistent with the perturbative computations in the field theory side \cite{c1}.

\subsubsection{Legendre transformation}
Firstly, we consider the loop with no dependence on the deformed $S^5$ and the only coordinate need to be replaced by its conjugate momentum is the radial coordinate $y$. From the Nambu-Goto action (\ref{ng}), it can be easily obtained as,
\be
P_y=\frac{\sqrt{1+C^2}R^2}{2\pi\alpha^\prime r^2}\frac{-f'f^2}{\sqrt{1+C^2+C^2f^2}}\frac{1}{\sqrt{f^4+f^2+f'^2}}.\label{py}
\ee
Near the boundary $y=\epsilon$ or $f=r/\epsilon$, we can evaluate $f'$ approximately from (\ref{f2}),
\be
\left(\frac{df}{d\psi}\right)^2\approx\frac{r^6}{\epsilon^6}\frac{1+C^2(1+f_0^2)}{C^2f_0^2(1+f_0^2)},
\ee
which indicates $f'^2\gg f^4\gg f^2$, thus we obtained from eq.~(\ref{py}),
\be
P_y\approx\frac{\sqrt{1+C^2}R^2}{2\pi\alpha^\prime C r \epsilon}.
\ee
So the boundary term is,
\be
-2\int^{L}_\epsilon dr(P_yy)|_{y=\epsilon}\approx-\frac{\sqrt{1+C^2}R^2}{\pi\alpha^\prime C}\log\frac{L}{\epsilon}.
\ee
Notice this cannot be used to cancel the leading $\log^2$ divergence found in the previous section. The computation of the Legendre transformation  for the case with a jump in deformed $S^5$ is similar and we arrive at the same conclusion.

\subsubsection{Geometric subtraction}
We may consider a geometric subtraction scheme which is performed by discarding two `flat' plane in the deformed AdS space with the metric,
\be
ds^2=R^2\sqrt{1+C^2}\left[\frac{dy^2+dr^2}{y^2+C^2(y^2+r^2)}+\frac{C^2(y^2dy^2+r^2dr^2+2yrdydr)}{y^2(y^2+C^2(y^2+r^2))}\right].
\ee
So the area to be subtracted is,
\bea
A_s&=&2\int dy dr \sqrt{G_{yy}G_{rr}-G_{ry}^2}\nonumber\\
&=&2R^2\sqrt{1+C^2}\int^{y_2}_{y_1} dy \int^{r_2}_{r_1} dr\frac{1}{y\sqrt{y^2+C^2(y^2+r^2)}}\nonumber\\
&\approx&\frac{2R^2\sqrt{1+C^2}}{C}\left(\log\frac{2C}{\sqrt{1+C^2}}\log\frac{y_2}{y_1}+\log r_2\log\frac{y_2}{y_1}-\frac{1}{2}\log^2 y_2+\frac{1}{2}\log^2y_1\right),
\eea
where $y_1,r_1\equiv \epsilon $ and $y_2,r_2\equiv L$ are the IR and UV cutoffs respectively. 
In another word, we have 
\bea A_s&\approx&\frac{2R^2\sqrt{1+C^2}}{C}\left(\log\frac{2C}{\sqrt{1+C^2}}\log\frac{L}{\epsilon}+\log L\log\frac{L}{\epsilon}-\frac{1}{2}\log^2 L+\frac{1}{2}\log^2\epsilon\right). \eea
From this result, one can see that
the leading $\log^2$ divergence can not be canceled using this geometric substraction. One can also see that the Legrendre transformation is not equivalent
with the geometric substraction, as we advertised previously.

\section{Anomalous dimension from cusped Wilson loop}
The anomalous dimension can be obtained by the vacuum expectation value of  a light-like Wilson loop with a cusp \cite{c3}. We will only consider the case without a jump in deformed $S^5$ at the cusp. The light-like system can be reached from the solution we have found by analytically continuing $f_0\rightarrow i f_0$ and taking $f_0$ to a fixed value which will be given later, thus the cusp angle $\Omega$ becomes $\pi+i\gamma$ ,
with \footnote{The real part of $\Omega$, which equals to $\pi$, comes from the residue at $z=f_0$.} \be
\gamma=P.P.\int^{+\infty}_{-\infty} dz \frac{f_0\sqrt{1-f_0^2}\sqrt{1+C^2-C^2f_0^2+C^2z^2}}{(z^2-f_0^2)\sqrt{1-f_0^2+z^2}\sqrt{z^2-2f_0^2+1+C^2(1-f_0^2)(1-f_0^2+z^2)}}.
\ee
The (renormalized) area $A$ now becomes
\bea
A&=&\sqrt{(1+C^2)(1+C^2-f_0^2C^2)}R^2\log\frac{L}{\epsilon}\int^{+\infty}_{-\infty} dz\nonumber\\
&&\left\{\frac{\sqrt{1+z^2-f_0^2}}{\sqrt{1+C^2-C^2f_0^2+C^2z^2}\sqrt{z^2-2f_0^2+1+C^2(1-f_0^2)(1-f_0^2+z^2)}}\right.\nonumber\\
&&\left.-\frac{1}{zC\sqrt{1+C^2-C^2f_0^2}}\right\}.
\eea
As discussed in the previous section, neither Legendre transformation nor geometric substraction can cancel the leading $\log^2$ divergence,
and to extract the anomalous dimension which comes from the coefficient of the logarithmic divergence, we subtract the leading divergence by hand
in the above expression.
In order to make the above two integrals real when $z\rightarrow 0$, 
we can choose  $f_0^2$ to satisfy
\be
f_0^2\le\frac{C^2+1-\sqrt{C^2+1}}{C^2}.
\ee
then we make the transformation
\be f_0^2=\frac{C^2+1-\sqrt{C^2+1+C^2\delta}}{C^2}, \ee
which gives  \be \delta=1-2f_0^2+C^2(1-f_0^2)^2,\ee and the integral can be expressed in terms of $\delta$ as
\be
\gamma=\int^{+\infty}_{-\infty} dz \frac{\sqrt{\frac{C^2+1-\sqrt{C^2+1+C^2\delta}}{C^2}}\sqrt{\frac{\sqrt{1+C^2+C^2\delta}-1}{C^2}}\sqrt{\sqrt{1+C^2+C^2\delta}+C^2z^2}}
{\frac{C^2z^2-C^2-1+\sqrt{C^2+1+C^2\delta}}{C^2}\sqrt{\frac{\sqrt{1+C^2+C^2\delta}-1}{C^2}+z^2}\sqrt{\sqrt{C^2+1+C^2\delta}z^2+\delta}}.
\ee
In order for the two edges of the cusped Wilson loops to be light-like, we need to take a limit such that $\gamma\to \infty$. This limit is given by $\delta\rightarrow 0$ (which obviously corresponds to $f_0^2\rightarrow \frac{C^2+1-\sqrt{C^2+1}}{C^2}$), one can see the largest contribution stems from the term $\sqrt{\sqrt{C^2+1+C^2\delta}z^2+\delta}$ around $z\approx 0$, i. e. $z\in (-\epsilon, \epsilon)$. When $\delta\ll \epsilon \ll 1$,we get
\be
\gamma\approx\frac{C}{\sqrt{C^2+1-\sqrt{C^2+1}}}\log\delta.
\ee
The same method can be applied to compute the area, which gives,
\be
A\approx-\frac{R^2(C^2+1)^{1/4}\sqrt{\sqrt{C^2+1}-1}}{C}\log\delta\log\frac{L}{\epsilon}.
\ee
So the 
cusp anomaly is,
\be
\bar{\Gamma}_{cusp}=-\frac{A}{2\pi\alpha^\prime|\gamma|\log\frac{L}{\epsilon}}=-\frac{R^2(1+C^2-\sqrt{1+C^2})}{2\pi\alpha^\prime C^2}.
\ee
In the $C\to 0$ limit, we have \be \bar{\Gamma}_{cusp}=-\frac{R^2}{4\pi\alpha^\prime}.\ee
By using the relation $R^2=\alpha^\prime \sqrt{\lambda}$ in the undeformed case with $\lambda$ the 't Hooft of the ${\cal N}=4$ super Yang-Mills theory,
we get \be\bar{\Gamma}_{cusp}=-\frac{\sqrt{\lambda}}{4\pi},\ee
which coincides with the results obtained in \cite{c3}.

\section{Conclusion and discussions}
$q$-deformed $AdS_5\times S^5$ is a quite interesting background of type IIB string theory. It is integrable, and its dual field theory
is still unclear. Probably it is dual to certain non-local field theory.  We hope various computations on the string theory side could give some
hints on the possible dual field theory. The study in this paper add one more example of such computations.  We studied the F-string theory solution
dual to  cusped Wilson loops in the field theory side. Both the case with a jump in deformed $S^5$ at the cusp and the case without such jump are studied.

The first interesting aspect we find for the cusped Wilson loops is the divergence behavior of the area of the F-string worldsheet before renormalization. In \cite{c4}, it is shown that the F-string dual to circular Wilson loop has finite area, before any
renormalization or Legendre transformation though the action of F-string dual to straight line is divergent \cite{Kameyama:2014vma}. The first result was explained by that
the deformation parameter $C$ provides a UV regularization \cite{c4}.
 For the Wilson loops with a cusp, the bare area goes like $\frac{\sqrt{1+C^2}R^2}{C}\log^2\frac{L}{\epsilon}-\frac{1}{2\pi}F(\Omega,\Theta,C)\log\frac{L}{\epsilon}$. Now the leading divergence scaling as $\log^2\frac{L}{\epsilon}$ which is not made finite due to the deformation, however it is less divergent than  $L/\epsilon$. In other words,
$q$-deformation softens the divergence while does not soften it into a finite term. This divergence can be removed neither by the Legendre transformation
nor via geometric substraction, and these two methods are no longer equivalent to each other.

Another feature of our solution is that the cusped anomalous dimension obtained from the solution without a jump in deformed $S^5$ can be smoothly connected
with the result in the undeformed case when we take the limit that the deformation parameter $C$ tends to zero. This is quite different from the
case for the spinning folded GKP-like string \cite{Kameyama:2014vma, FrolovRoiban}. For the GKP string, the relation $E-S\sim f(\lambda)\log S$ for large $S$ was  destroyed
by the deformation. This makes us not be able to extract the anomalous dimension from the GKP string side and compare with the results given here from
the cusped Wilson loops. The equivalent of these two approach for the undeformed case is broken down by the deformation partly because the background
has much smaller isometry group after deformation.

For the undeformed case, the solution dual to the cusped Wilson loop with two light-like edges in \cite{c3} was found to actually  have four cusps using the embedding coordinations \cite{Alday:2007hr}. Such observation made this solution to play a key role in the holographic computations of four-gluon planar amplitudes at strong coupling. It should be interesting to  try to  embed the deformed $AdS_5$ into a higher-dimensional spacetime and study the geometry of the minimal surface dual to the cusped Wilson loop from this point of view.

It should be interesting to compute the holographic entanglement entropy \cite{Ryu:2006bv, Ryu:2006ef, Hubeny:2007xt}  from this background to investigate whether the area law \cite{Bombelli:1986rw, Srednicki:1993im} of the entanglement entropy is lost or not, since probably the dual field theory is a non-local one. However to perform this computation, we need to know the metric in the Einstein frame. Since the metric
in the string frame is known, we need to know the dilaton field. Some progresses were made in \cite{Lunin:2014tsa, ABF}, however complete solution is still unknown.
It should be valuable to find consistent solution including dilaton and Ramond-Ramond field and compute the holographic entanglement entropy. We hope we could work on this point in the near future.

\section*{Acknowledgments}
We would like to thank  Xiao-Ning Wu, Gang Yang, Jia-Ju Zhang and Fen Zuo for very helpful discussions. J.~W. would like to thank Institute of Astronomy and Space Science, Sun Yat-Sen University and School of Physics, Huazhong University of Science and Technology  for warm hospitality during a recent visit.  This work was in part supported by NSFC Grant No.~11222549 (N.~B. and J.~W.), and No.~11275207 (H.~C.). J.~W. also gratefully acknowledges the support of K.~C.~Wong Education Foundation and Youth Innovation Promotion Association of CAS.

\end{document}